%% file: main.tex
\documentclass[
    aps,
    pre,
    onecolumn,
    10pt,
    notitlepage,
    superscriptaddress,
    longbibliography,
    nofootinbib
]{revtex4-2}
\usepackage[english]{babel}

\usepackage{amsmath}
\usepackage{amsfonts}
\usepackage{amssymb}
\usepackage{cancel}

\usepackage{graphicx}
\usepackage{tikzfeynman}

\usepackage{booktabs}
\usepackage{array}
\usepackage{makecell}
\usepackage{multirow}
\usepackage{xcolor}
\usepackage{comment}
\usepackage{todonotes}
\usepackage{float}

\usepackage[colorlinks=true,
            citecolor=blue,
            linkcolor=blue,
            urlcolor=blue]{hyperref}
\usepackage{cleveref}

\begin{document}

\title{Dynamic critical exponent of the Yang--Lee edge singularity at three loops}

\author{Loran Ts. Adzhemyan}
\affiliation{
Saint Petersburg State University,
7/9 Universitetskaya nab.,
St. Petersburg 199034,
Russian Federation
}
\affiliation{
Bogolyubov Laboratory of Theoretical Physics,
Joint Institute for Nuclear Research,
Dubna 141980,
Russian Federation
}

\author{Diana A. Davletbaeva}
\affiliation{
Saint Petersburg State University,
7/9 Universitetskaya nab.,
St. Petersburg 199034,
Russian Federation
}
\affiliation{{Petersburg Nuclear Physics Institute named by B.P.Konstantinov of NRC ``Kurchatov Institute'', mkr. Orlova roshcha 1, Gatchina, Leningradskaya oblast, 188300, Russian Federation}}

\author{Daniil A. Evdokimov}
\affiliation{
Saint Petersburg State University,
7/9 Universitetskaya nab.,
St. Petersburg 199034,
Russian Federation
}
\affiliation{
Bogolyubov Laboratory of Theoretical Physics,
Joint Institute for Nuclear Research,
Dubna 141980,
Russian Federation
}

\author{Mikhail V. Kompaniets}
\email{m.kompaniets@spbu.ru}
\affiliation{
Saint Petersburg State University,
7/9 Universitetskaya nab.,
St. Petersburg 199034,
Russian Federation
}
\affiliation{
Bogolyubov Laboratory of Theoretical Physics,
Joint Institute for Nuclear Research,
Dubna 141980,
Russian Federation
}

\begin{abstract}
We compute the dynamic critical exponent $z$ of the Yang--Lee edge singularity in relaxational dynamics using perturbative renormalization-group methods to the three-loop order. The calculation combines diagram reduction, analytic two-loop evaluation via parametric integration with hyperlogarithms, and numerical evaluation of three-loop integrals using the Sector Decomposition method. The perturbative series obtained is resummed using Padé and Padé-Borel-Leroy methods to produce estimates of $z$ in various spatial dimensions. The resulting values are consistent with previous perturbative and functional renormalization-group calculations.
\end{abstract}

\maketitle

\section{Introduction}
The Yang--Lee edge singularity (YLES) problem, originally formulated in \cite{LeeYang1952_1,LeeYang1952_2}, has since become a classic subject in statistical physics. A decisive contribution was made by Kortman and Griffiths \cite{KorthGrif_1971} and Fisher \cite{Fisher1978}, who uncovered its close connection to critical phenomena, namely the emergence of a divergent correlation length and universal critical exponents governing scaling behavior. This connection enabled the application of the then newly developed renormalization group (RG) theory and perturbative $\varepsilon$-expansion techniques. Fisher demonstrated that the scaling behavior of the YLES is described by a scalar field theory with a cubic $\phi^3$ interaction and a purely imaginary coupling constant. While the standard $\phi^3$ theory with a real coupling is unstable due to the absence of a real infrared (IR) attractive fixed point, the theory with an imaginary coupling possesses a real-valued IR fixed point.
The field-theoretic RG approach has allowed for a detailed investigation of both static and dynamic properties of the YLES. Various methods have been used to estimate static critical exponents, most notably perturbative RG \cite{Fisher1978, Bonfirm_PRG_YLES, Janssen1981, Gracey2015_PRG_YLES, BGKS_phi321, Schnetz_6l_phi3, Gracey_sixloop_YLES}, functional RG \cite{Zhong_FRG,An_FRG_YLES,zambelli_FRG_YLES,Rennecke_FRG_YLES,Johnson_FRG_YLES}, and conformal bootstrap \cite{Gliozzi_conformYLES,Hikami_conformYLES1,Hikami_conformYLES2}. In contrast, the dynamic properties of the YLES remain much less explored. The central quantity characterizing dynamics is the dynamic critical exponent $z$, which determines the critical slowing down of the relaxation time $\tau$
\begin{equation}
\tau \propto \xi^z\,,
\end{equation}
where $\xi$ denotes the diverging correlation length. To date, available results are essentially limited to the two-loop calculation reported in 1981 \cite{Janssen1981} and a more modern functional RG estimate \cite{Zhong_FRG}. The scarcity of results is primarily due to the much higher technical complexity of dynamic RG calculations compared to their static counterparts. In addition, direct observation of dynamic scaling behavior in the YLES, either experimentally or via numerical simulations, is extremely challenging. Most modern empirical studies focus instead on the quantum YLES \cite{Yin_sim_YLES, Zhai_exp_YLES, Gao_exp_YLES, Sun_exp_YLES}, where both static and dynamic scaling have been confirmed. However, to our knowledge, no experimental estimates of dynamic critical {exponents} for the classical YLES have been reported.

Renewed interest in the dynamics of the classical YLES has been stimulated by a series of works \cite{Zhong1995, Zhong2005, Zhong2011, Zhong2017, Zhong2024, Zhang_Zhong_2025}  demonstrating the emergence of universal scaling behavior, traditionally associated with continuous phase transitions, also in first-order phase transitions (FOPTs). In this approach, the dynamics of generic FOPTs driven by an external field in the $\phi^4$ model below its critical point is described by an instability fixed point of the corresponding $\phi^3$ theory. Consequently, the large-scale behavior of the scalar $\phi^3$ theory describing FOPTs belongs to the same universality class as the YLES. Since this scenario relies on the nontrivial assumption that an imaginary fixed point can control the real infrared asymptotic scaling behavior (see supporting arguments in \cite{Zhong_imag_fp}), more accurate determinations of critical exponents are essential to further substantiate this framework \cite{Zhong2024}. While the static critical exponents of the $\phi^3$ (YLES) theory are already known to high perturbative orders \cite{Gracey_sixloop_YLES}, obtaining a more precise estimate of the dynamic exponent $z$ would be highly desirable.

The primary goal of this work is to obtain a three-loop estimate of the dynamic critical exponent $z$ for the YLES with purely relaxational dynamics, corresponding to model A in the Hohenberg-Halperin classification \cite{Halperin77}. As a preliminary step, we applied a diagram reduction technique \cite{AIKV_4lSD17} to decrease the number of Feynman diagrams that require explicit evaluation. The calculations were performed using two approaches: the numerical Sector Decomposition method \cite{BH_sd04} and the analytic technique of parametric integration with hyperlogarithms \cite{Brown08, Panzer15}. The latter is a well-established tool for multiloop calculations and has recently been successfully applied to models of critical dynamics, including stochastic turbulence and model A of the $\phi^4$ theory \cite{AEKTurb2024, ADEK26}.  {Applying this technique to dynamic models is, however, more challenging because the specific structure of the corresponding Feynman integrals may obstruct linear reducibility \cite{Panzer_lr14}.}
{These developments motivate the application of the same approach to the dynamic $\phi^3$ theory describing the YLES.} Although we were only able to reproduce the previously known analytic two-loop result \cite{Janssen1981} using the hyperlogarithmic approach, this analysis clarified several technical aspects of its implementation. The main result of this work is a numerical three-loop calculation obtained by Sector Decomposition adapted for the evaluation of dynamic diagrams \cite{AIKV_4lSD17, SD_gitlab}, followed by the resummation of the computed three-loop contributions. Unlike earlier studies, where the limited perturbative order restricted the analysis to Padé resummation \cite{Zhong2017}, the availability of the third-order contribution allows us to employ more advanced techniques, including Padé-Borel resummation and the KP17 method \cite{KP17}.

The paper is organized as follows. In Section~\ref{sec2: Renormalization of the model}, we briefly review the renormalization of the dynamic generalization of the scalar cubic model related to the YLES. In Section~\ref{sec:3_Diagrammatic technique}, we introduce the dynamic Feynman diagrams of the model and their representation in terms of the so-called time versions. In Section~\ref{sec:4_Diagram_reduction}, we discuss a diagram-reduction procedure that decreases the number of diagrams to be evaluated. Section~\ref{sec:5_Two-loop_parametric_integration} is devoted to deriving the analytic two-loop result via parametric integration; in particular, we describe in detail the regularization of the integrals required by this method. In Section~\ref{sec:6_Three-loop_RG_functions}, we use a numerical Sector Decomposition approach to compute the three-loop RG functions and the exponent $z$. Finally, in Section~\ref{sec:7_Resummation}, we resum the perturbative series using Padé, Padé-Borel, and the KP17 method.

\section{Renormalization of the model}
\label{sec2: Renormalization of the model}
The action associated with the Yang--Lee singularity problem has the form of the static $\phi^3$ theory with an imaginary coupling constant $i {g_0}$:
\begin{equation}
S^{st}(\psi_0) = -\int d^d x \left[\frac{1}{2}(\partial \psi_0)^2 + \frac{1}{2}m_0^2\psi_0^2 + i\frac{g_0}{3!} \psi_{0}^3\right]\,.
\label{statics}
\end{equation}
 
 The dynamic generalization of the static theory with a nonconserved order parameter (model A) is based on the Langevin-type formulation
\begin{equation}
S(\psi_0,\psi'_0)= \lambda_0 \psi'_0\psi'_0+\psi'_0\left[-\partial_t\psi_0+\lambda_0\frac{\delta S^{st}(\psi_0)}{\delta \psi_0}\right]\,,
\end{equation}
where $\psi'_0$ is an auxiliary field and $\lambda_0$ is the Onsager coefficient. The explicit form of the unrenormalized action for the dynamic generalization of \eqref{statics} is given by
\begin{equation}
\label{base_action}
S(\psi_0,\psi'_0) = \lambda_0 \psi'_0 \psi'_0 + \psi'_0\left[-\partial_t \psi_0 + \lambda_0 (\partial^2 \psi_0 - m_0^2 \psi_0 - \frac{1}{2}i g_0 \psi_0^2)\right]\,.
\end{equation}
This model is multiplicatively renormalizable; in spatial dimension $d=6-2\varepsilon$ the renormalized action reads
\begin{equation}
\label{renorm_action}
S_R (\psi,\psi')= Z_1 \lambda \psi' \psi' + \psi' \left [-Z_2 \partial_t \psi + \lambda (Z_3 \partial^2 \psi - Z_4 m^2 \psi - \frac{1}{2} Z_5 i g \mu^{\varepsilon} \psi^2)\right],
\end{equation}
where $\mu$ is the renormalization mass, and the renormalized fields and parameters are related to the bare ones by
\begin{equation}
\label{renorm_params}
\lambda_0 = \lambda Z_{\lambda}, \quad m_0 = m Z_m, \quad g_0 = g \mu^{\varepsilon} Z_g, \quad \psi_0 = \psi Z_{\psi}, \quad \psi'_0 = \psi' Z_{\psi'}.
\end{equation}
The renormalization constants \( Z_i \) are associated with the renormalization constants from \eqref{renorm_params} by the following relations:
\begin{equation}
Z_1 = Z_\lambda Z_{\psi'}^2, \quad Z_2 = Z_{\psi'}Z_\psi , \quad Z_3 = Z_{\psi'} Z_\lambda Z_\psi,
\label{stat_dyn_equiv}
\end{equation}
\begin{equation}
Z_4 = Z_{\psi'} Z_\lambda Z_m^2 Z_\psi, \quad Z_5 = Z_{\psi'} Z_\lambda Z_g Z_\psi^2.
\end{equation}

Since the renormalization constants \( Z_\psi \), \( Z_m \), and \( Z_g \) coincide with their static counterparts \cite{Vasiliev_book}
\begin{equation}
\label{dyn_stat_eq}
Z_\psi = (Z_\psi)_{st}, \quad Z_m = (Z_m)_{st}, \quad Z_g = (Z_g)_{st},
\end{equation}
and because the relation \( Z_{\psi'} Z_\lambda = Z_\psi \) holds \cite{Vasiliev_book}, it follows that $Z_1=Z_2$. We are interested only in \( Z_\lambda \), which characterizes the dynamic model:
\begin{equation}
Z_\lambda = Z_1 Z_{\psi'}^{-2} = Z_1^{-1} Z_\psi^2 = Z_2^{-1} Z_\psi^2.
\end{equation}
For practical purposes, it is convenient to evaluate it using the renormalization constant \( Z_1 \) which we determine from diagrams of the one-particle irreducible Green function \( \Gamma_{\psi' \psi'} = \langle \psi' \psi' \rangle_{1-\text{irr}} / (2 \lambda) \) considered at zero external frequency \( \omega =0\) and \( m = 0 \). Henceforth, we use the redefined coupling constant \( u = g^2\, S_d / (2\pi)^d \), where \( S_d = 2\pi^{d/2} / \Gamma(d/2) \) is the area of the \( d \)-dimensional sphere of unit radius. The perturbative expansion of the renormalized Green function reads
\begin{equation}
\label{gamma_expansion}
\Gamma^R_{\psi' \psi'} \big|_{\omega=0, m=0} = Z_1 \left( 1 + \sum_{l=1}^{\infty} \left(-u \, Z_u \, \left(\frac{\mu^2}{p^2}\right)^{\varepsilon}\right)^l A^{(l)} \right),
\end{equation}
where $A^{(l)}$ are diagrammatic contributions of order $l$ of $\Gamma_{\psi' \psi'}|_{\omega=0, m=0}$.

The renormalization constant $Z_1$ in the three-loop approximation in the minimal subtraction (MS) scheme has the form
\begin{equation}
\label{MS}
  Z_1(\varepsilon,u) = 1
    + u \Bigl(\frac{b_{11}}{\varepsilon}\Bigr)
    + u^2 \Bigl(\frac{b_{22}}{\varepsilon^2} + \frac{b_{21}}{\varepsilon}\Bigr)
    +  u^3 \Bigl(\frac{b_{33}}{\varepsilon^3} + \frac{b_{32}}{\varepsilon^2} + \frac{b_{31}}{\varepsilon}\Bigr) + \mathcal{O}(u^4).
\end{equation}
The coefficients $b_{ij}$ are fixed by the requirement that all poles in \eqref{gamma_expansion} are canceled. They are then expressed in terms of diagrammatic contributions $A^{(l)}$ taking the form
\begin{equation}
\label{A_i}
A^{(1)} = \frac{A_{11}}{\varepsilon} + A_{10} + A_{1,-1}\,\varepsilon\,,
\qquad 
A^{(2)} = \frac{A_{22}}{\varepsilon^2} + \frac{A_{21}}{\varepsilon} + A_{20},
\qquad
A^{(3)} = \frac{A_{33}}{\varepsilon^3} + \frac{A_{32}}{\varepsilon^2} + \frac{A_{31}}{\varepsilon}\,.
\qquad
\end{equation}
The Green function also involves the renormalization constant for the coupling constant, known from the static theory \cite{Vasiliev_book}, of which only the first two orders in $u$ are required:
\begin{equation}
     Z_u(\varepsilon,u) = 1 + u \left(\frac{c_{11}}{\varepsilon}\right)
    + u^2 \Bigl(\frac{c_{22}}{\varepsilon^2} + \frac{c_{21}}{\varepsilon}\Bigr)+ \mathcal{O}(u^3)\,,
\end{equation}
\begin{equation}
\label{c_i}
     c_{11} = \frac{3}{4}\,,\,\,\,\,\,\,\,\, c_{22} =  \frac{9}{16}\,,\,\,\,\,\,\,\,\, c_{21} = -\frac{125}{288}\,.
\end{equation}
Using \eqref{gamma_expansion}, \eqref{MS}, \eqref{A_i}, and \eqref{c_i}, one can obtain the renormalization constant $Z_1$ in the three-loop approximation.

\section{Diagrammatic technique} \label{sec:3_Diagrammatic technique}
The propagators corresponding to the model \eqref{base_action} can be written in the $(k,t)$ representation as follows:
\begin{equation}
\langle \psi(t_1) \psi (t_2) \rangle = \frac{1}{k^2} e^{-\lambda k^2|t_1-t_2|} =\vcenter{\hbox{ 
\begin{tikzpicture}[node distance=1.5cm]
\coordinate (v1); 
\coordinate[right=of v1] (v2);
\arcl{v1}{v2}{0};
%\draw (v1) -- ($ (v1)!.15!180:(v3) $);
\end{tikzpicture}}},
\label{prop_psipsi}
\end{equation}

\begin{equation}
    \hspace{1.1cm} \langle \psi(t_1) \psi' (t_2) \rangle = \theta(t_1-t_2) e^{-\lambda k^2(t_1-t_2)} = \vcenter{\hbox{
    \begin{tikzpicture}[node distance=1.5cm]
    \coordinate (v1); 
    \coordinate[right=of v1] (v2);
    \coordinate(e2) at ($ (v1)!.15!(v2) $);
    \coordinate[above=of e2] (sv1);
    \coordinate[below=of e2] (sv2);
    \arcl{v1}{v2}{0};
    \draw ($ (v2)!.15!0:(sv1) $) -- ($ (v2)!.15!0:(sv2) $);
    \end{tikzpicture}}},
\label{prop_psi'psi}
\end{equation}

\begin{equation}
\hspace{-3.95cm} \langle \psi'(t_1) \psi'(t_2) \rangle = 0 .
\label{prop_psi'psi'}
\end{equation}

The prefactor of the diagram is determined by its symmetry and the additional factor $1/2$ in the definition of the one-particle irreducible Green function $\Gamma_{\psi' \psi'} = \langle \psi' \psi' \rangle_{1-\text{irr}} / 2\lambda.$ Because propagators \eqref{prop_psipsi} and \eqref{prop_psi'psi} depend on time exponentially, integrations over times become trivial, unlike the $(k,\omega)$ representation, which requires contour integration. Since the renormalization of the model is considered at $\omega=0$, it is convenient to employ the time-versions technique, which allows one to represent the result of time integrations in a compact diagrammatic form suitable for further manipulations \cite{Vasiliev_book}. Being convenient for automation, it has been actively used in a series of multiloop studies of dynamic models \cite{AIKV_4lSD17, AEKTurb2024, ADEK26, AVKK_Hmodel99, ADHIK_multi16, AEHIKKZ_5l_2_22}. Let us illustrate this with a two-loop example:
\begin{equation}
 \label{3l_tvs}
    \includegraphics[width=1\textwidth]{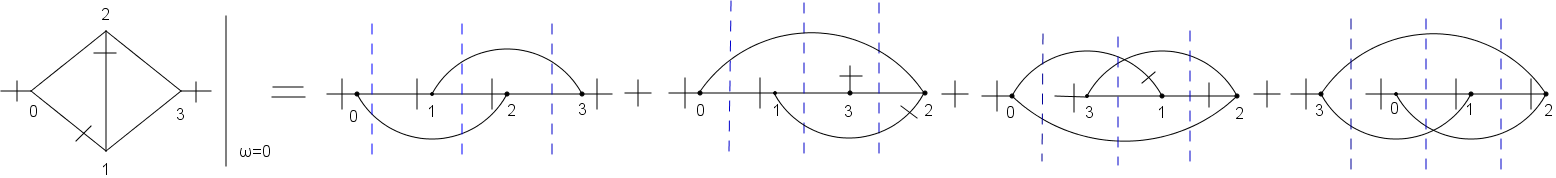} 
\end{equation}
The numbers $0,1,2,3$ represent the times $t_0,t_1,t_2,t_3$ associated with each vertex. The presence of the $\theta$-function in the propagator \eqref{prop_psi'psi} leads to a sum over time versions, each defined by an ordered sequence of times. Thus, each dynamic diagram can be represented as a sum of its time versions. This is the reason why the number of diagrams in dynamic models significantly exceeds the number of diagrams in static models.

The result of the time integrations can be schematically shown by dotted lines representing the so-called ``time cuts''. We will denote the square of the momentum flowing through a given line as its ``energy'' $E(k)=k^2$. Then, we associate a factor $1/E_{k_i}$ with each solid line $\langle \psi \psi \rangle$, a factor $1$ with each solid line with dash $\langle \psi \psi' \rangle$ and a factor $1/\sum_i E_{k_i}$ with each dotted line. The last sum goes over all energies of the diagram  crossed by the dotted line, where $k_i$ are the momenta of the corresponding lines. For example, the first time version of \eqref{3l_tvs} is expressed as follows

\begin{equation}
\label{tv}
\raisebox{-0.4\height}{\includegraphics[width=0.2\textwidth]{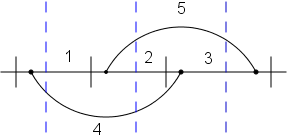}} \sim \frac{1}{E_3 E_4 E_5} \cdot \frac{1}{E_1+E_4} \cdot \frac{1}{E_2 + E_4 + E_5} \cdot \frac{1}{E_3+E_5}
\end{equation}
The remaining time versions can be expressed in a similar manner. Note that here the numbers label the lines, whereas in \eqref{3l_tvs} they label the vertices. The appearance of time-cut propagators, which are specific to dynamic models, significantly complicates the evaluation of diagrams compared to the static case.

\section{Diagram reduction}
\label{sec:4_Diagram_reduction}
One of the reasons why progress in models of critical dynamics lags behind their static counterparts is the much larger number of dynamic diagrams originating from the time dependence of propagators. In \cite{AIKV_4lSD17}, the authors proposed an algorithm for dynamic diagram reduction that significantly decreases their number. This algorithm is based on the equality of certain renormalization constants in statics and dynamics, which can be formulated as the following rule:
\begin{equation}
    \label{reduction_rules}
    \vcenter{\hbox{\input{tikz_pics/rule4}}} \quad + \quad \vcenter{\hbox{\input{tikz_pics/rule3}}} \quad + \quad \ldots \quad + \quad \vcenter{\hbox{\input{tikz_pics/rule2}}} \quad + \quad \vcenter{\hbox{\input{tikz_pics/rule1}}} \quad = \quad \vcenter{\hbox{\input{tikz_pics/rule5}}}
\end{equation}
Here, circles denote an arbitrary subgraph, which is identical in each graph, while there may be $k_1 \geq 2$ lines to the left of the subgraph and $k_2 \geq 0$ lines to the right. An iterative application of various forms of this rule to the original diagrams allows simplifying their structure (by eliminating certain time cuts) and reducing their total number (see \cite{AIKV_4lSD17} for details)\footnote{In \cite{ADEK26}, a more advanced reduction scheme was proposed, specifically aimed at eliminating dynamic subgraphs. It substantially reduces the number of subtractions required within the hyperlogarithms integration framework. However, in contrast to the original scheme, it has not yet been automated, and due to the large number of three-loop diagrams in the model under consideration, we do not apply it here.}.

\begin{figure}[t]
    \centering
    \includegraphics[width=0.8\textwidth]{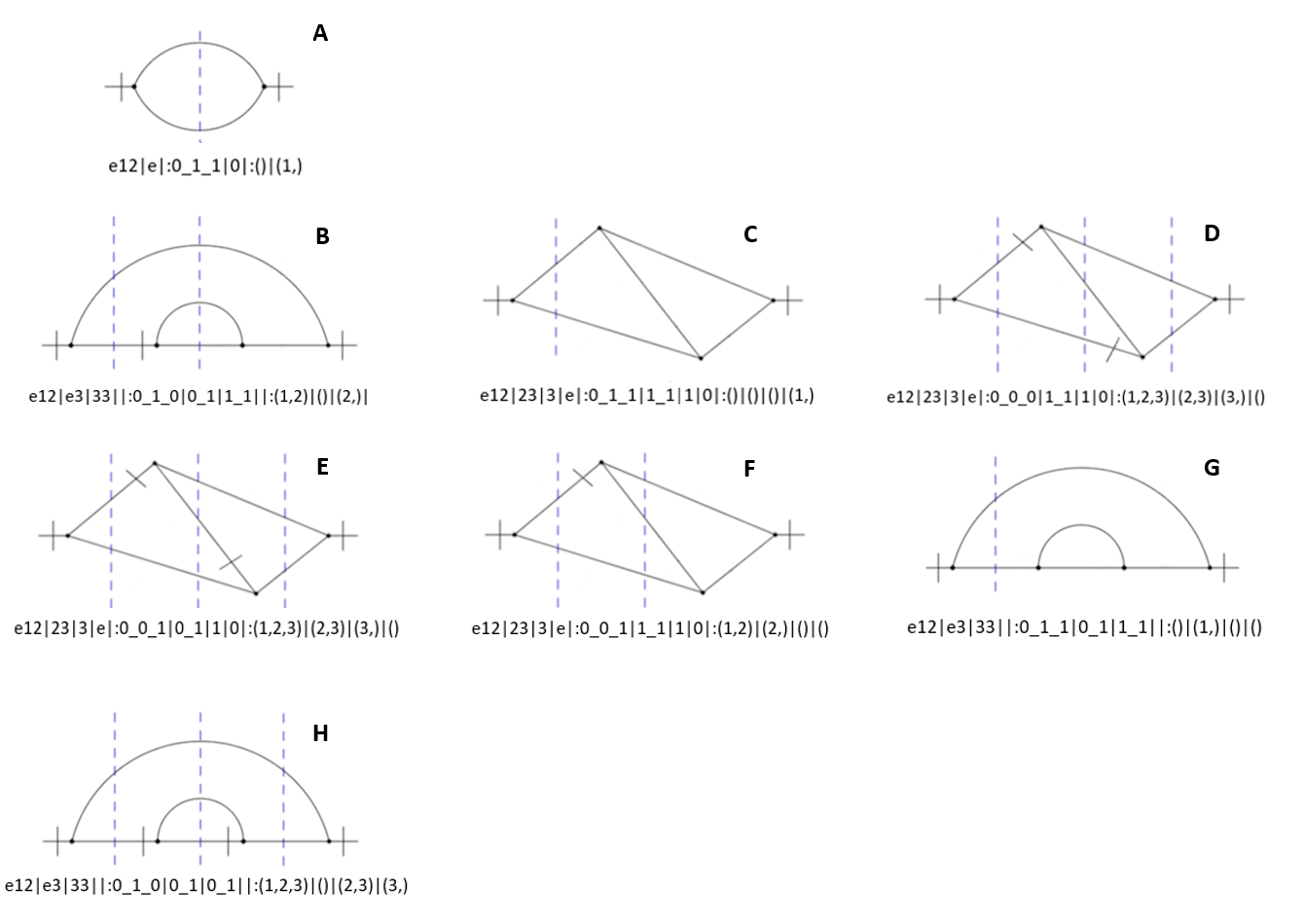} 
    \caption{The list of all one- and two-loop time versions with their corresponding Nickel-like indices after performing the diagram reduction. The assignment of indices to dynamic diagrams with time cuts is described in \cite{ADEK26}.}
    \label{2loop_reduced}
\end{figure}

Diagram reduction was performed for both the two- and three-loop diagrams. The reduced two-loop time versions are shown in Fig.~\ref{2loop_reduced}, where their number falls from $28$ to $7$. The indices of the reduced three-loop time versions, together with the corresponding results, are collected in the Supplemental Material~\cite{SM}; the number of three-loop versions is reduced from $2044$ to $221$.

\section{Two-loop parametric integration}
\label{sec:5_Two-loop_parametric_integration}
A two-loop analytic calculation is performed using parametric integration via hyperlogarithms. The method was originally proposed by F.~Brown~\cite{Brown08} and subsequently automated by E.~Panzer in the Maple-based package \textit{HyperInt}~\cite{Panzer15, panzerPhD15}. The main requirement for this approach is the \textit{linear reducibility} of an integral, which ensures that the integrand can be expressed in terms of hyperlogarithms at each step of the integration procedure. Many diagrams arising in different QFT models satisfy this condition (e.g.\ in $\phi^4$ theory, the first non-reducible diagram appears only at six-loop order); as a result, this approach has become a standard tool in analytic multiloop calculations in recent years and has been applied to a wide range of QFT models.

Applications of parametric integration to models of critical dynamics have been reported in \cite{AEKTurb2024, ADEK26}. In the latter, the authors analytically studied the $\phi^4$-based model A in the four-loop approximation. Given the similarity of the models (dynamic generalizations of the $\phi^3$ and $\phi^4$ models), we outline here only the main aspects of the computations and refer the reader to \cite{ADEK26} for details.

The parametric integration method applies only to finite integrals. Therefore, it requires a regularization procedure that eliminates superficial and subgraph divergences. The regularized integral is then evaluated, and the poles in $\varepsilon$ that contribute to the renormalization constants are recovered. Regularization is not unique and is typically chosen to facilitate automation. The general prescription is to construct the finite integral from the initial divergent diagram by introducing subtraction terms chosen so that they are simpler to evaluate. Following \cite{ADEK26}, we employ a BPHZ-like scheme \cite{Brown13} with $\mathcal{R'}$-operation to subtract divergences, adapted to dynamic models.

\noindent The convergent,  renormalized integral can be constructed in the form
\begin{equation}
    \label{renorm_int}
    G^R = \partial_p \mathcal{R'} G \bigg|_{p=\mu},
\end{equation}
where $G$ is the diagram under consideration and $p$ is an external momentum. Since the Green function \eqref{gamma_expansion} depends only on the dimensionless ratio $\mu/p$, one can set $p=1$ by measuring the momentum $p$ in units of $\mu$. $\mathcal{R'}$ is Bogolubov's operation, which recursively subtracts all subgraph divergences, while the remaining superficial divergence is eliminated by differentiating with respect to $p$. $\mathcal{R'}$-operation is defined by Zimmermann's forest formula
\begin{equation}
    \label{R'}
    \mathcal{R'} G = \prod_{i} (1-K_i)G\,\,,
\end{equation}
where the product runs over all divergent subgraphs of the diagram $G$. Its explicit form is determined by the subtraction operation $\mathcal{K}$, which generally depends on the choice of the renormalization scheme. 

In the aforementioned BPHZ approach, the counterterms are defined by the expansion around the renormalization point, i.e., certain values of external momentum and frequency, which we choose to be $p=1$ and $\omega=0$. The diagram reduction described in Section~\ref{sec:4_Diagram_reduction} leads to the presence of not only dynamic subgraphs but also static ones. For the latter, the renormalization point $p=1$ is chosen. In cases with multiple external momenta (e.g. a three-point graph), the infrared rearrangement (IRR) technique can be employed. This allows one to set some of the external momenta to zero in logarithmically divergent diagrams under the $\mathcal{K}\mathcal{R'}$-operation, for both static and dynamic subgraphs, without introducing infrared divergences. In the evaluation of all two-loop diagrams in Fig.~\ref{2loop_reduced}, IRR is applied to treat static and dynamic triangle subgraphs in diagrams C, D, and E. Following \cite{ADEK26}, for general logarithmically divergent subgraphs we define the corresponding $\mathcal{K}_{\text{log}}$-operation as follows:
\begin{equation}
\mathcal{K}_{\text{log}} \, G := 
\begin{cases}
\left. G_{\text{IRR}} \right|_{p = 1} & \text{if } G \text{ is static }, \\
\left. G_{\text{IRR}} \right|_{p = 1, \omega = 0} & \text{if } G \text{ is dynamic }.
\end{cases}
\end{equation}

As an example, let us consider the renormalization of diagram E. It has one subgraph requiring subtraction: a dynamic log-divergent triangle with two time cuts:
\begin{align}
     \partial_p \mathcal{R'} \,\, \raisebox{-0.4\height}{\includegraphics[width=0.2\textwidth]{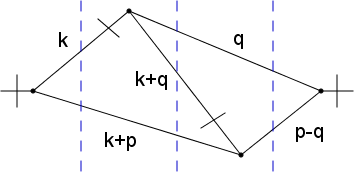}} \left.\vphantom{\rule{0pt}{20pt}}\right|_{p=1} \, & = \,\, \partial_p \left[ \raisebox{-0.4\height}{\includegraphics[width=0.2\textwidth]{pics/diag2_main.png}}\,\, - \,\, \mathcal{K}_{\text{log}} \raisebox{-0.4\height}{\includegraphics[width=0.16\textwidth]{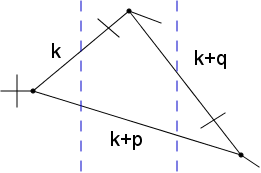}} \times \raisebox{-0.43\height}{\includegraphics[width=0.14\textwidth]{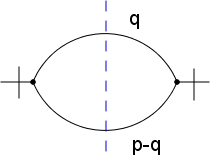}}\right] \left.\vphantom{\rule{0pt}{20pt}}\right|_{p=1}\notag \\
    & =  \,\,\, \partial_p \raisebox{-0.4\height}{\includegraphics[width=0.2\textwidth]{pics/diag2_main.png}} \left.\vphantom{\rule{0pt}{20pt}}\right|_{p=1} - \,\,\, \raisebox{-0.4\height}{\includegraphics[width=0.16\textwidth]{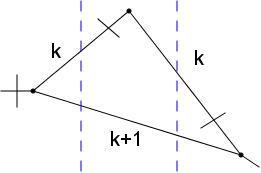}} \times \,\,\partial_p \raisebox{-0.43\height}{\includegraphics[width=0.14\textwidth]{pics/diag2_sub_term2.png}} \left.\vphantom{\rule{0pt}{20pt}}\right|_{p=1}\,\,,
\end{align}
Here, IRR is used to nullify one of the external momenta $q$ in the triangle. Because two-point graphs have a trivial momentum dependence, differentiation with respect to $p$ yields a factor ${-2l\varepsilon}$, where l is the number of loops. The original unrenormalized diagram can be reconstructed at $p=1$ as follows:
\begin{equation}
    \label{diag2_unrenorm}
    \raisebox{-0.4\height}{\includegraphics[width=0.2\textwidth]{pics/diag2_main.png}} \left.\vphantom{\rule{0pt}{20pt}}\right|_{p=1} \, = \,\,\, -\frac{1}{4\varepsilon} \partial_p \mathcal{R'} \raisebox{-0.4\height}{\includegraphics[width=0.2\textwidth]{pics/diag2_main.png}} \left.\vphantom{\rule{0pt}{20pt}}\right|_{p=1} \, + \,\,\, \frac{1}{2} \raisebox{-0.4\height}{\includegraphics[width=0.16\textwidth]{pics/diag2_sub_term1.png}} \times \,\,\raisebox{-0.43\height}{\includegraphics[width=0.14\textwidth]{pics/diag2_sub_term2.png}} \left.\vphantom{\rule{0pt}{20pt}}\right|_{p=1}\,\,.
\end{equation}
The first term on the right-hand side is convergent and can be evaluated using the hyperlogarithm method. The diagrams in the second term are free of subdivergences and contain only a superficial divergence (the only pole is explicitly isolated in Euler $\Gamma$-function after Feynman parametrization); hence, they are suitable for direct calculation as well. Combining these contributions yields the two pole terms for diagram E. The renormalization of diagrams B, C, D, and G, which also contain logarithmically divergent subgraphs, is carried out in the same manner.

The only diagram containing a quadratically divergent subgraph is diagram H; the subgraph is a dynamic loop with a cut. Since quadratically divergent subgraphs were not encountered in \cite{ADEK26}, we consider this case in greater detail. For static theories, the prescription for treating quadratic subgraphs within the BPHZ scheme is straightforward: subtract an additional term that contains a derivative in the expansion in external momenta \cite{Brown13}. However, the presence of an external frequency in dynamic diagrams precludes a direct generalization of this rule. A possible solution lies within the basic concept of regularization stated above: to construct, from the original integral, a convergent one by introducing sufficiently simple subtraction terms. For the considered diagram H, this strategy can be implemented as follows. Let us define the function
\begin{equation}
        I_H(a,b) =\int d^{d}k \,\,d^{d}q\, \frac{1}{(p+k)^2}\frac{1}{q^2}\frac{1}{k^2}\,\frac{1}{k^2+(p+k)^2}\frac{1}{q^2+(b\,k+q)^2+a\,(p+k)^2}\,\frac{1}{k^2+(p+k)^2}\,\,,
\end{equation}
such that $I_H(a=1,b=1)$ is an integral corresponding to diagram H\footnote{The factor $S_d/(2\pi)^d$  for each momentum integration is taken into account in the definition of the coupling constant $u=g^2{S_d}/{(2\pi)^d}$.}:
\begin{equation}
    {\includegraphics[width=0.22\textwidth]{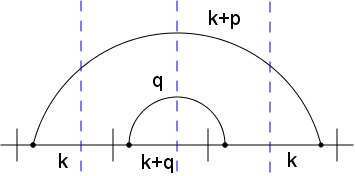}}
\end{equation}
Additional factors $a$ and $b$ are introduced to facilitate the subtraction procedure. One can verify that a combination
\begin{equation}
    \label{H_diag_renorm_nodiag}
    \tilde{\mathcal{R}'}I_H(1,1) = I_H(1,1)-I_H(0,1)-I_H(1,0)+I_H(0,0)
\end{equation}
is free of subdivergences and contains only a logarithmic superficial divergence. This cancellation of subdivergences can be illustrated diagrammatically as follows:
\begin{align}
    \label{H_diag_renorm}
     \tilde{\mathcal{R}'}\,\, & \raisebox{-0.4\height}{\includegraphics[width=0.2\textwidth]{pics/diag_H_with_mom.png}} = \notag \\  &   \raisebox{-0.4\height}{\includegraphics[width=0.20\textwidth]{pics/diag_H_with_mom.png}} - \raisebox{-0.4\height}{\includegraphics[width=0.20\textwidth]{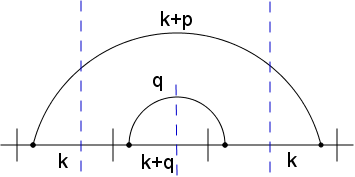}} - \raisebox{-0.4\height}{\includegraphics[width=0.20\textwidth]{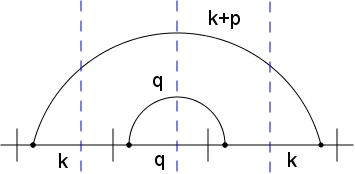}} + \raisebox{-0.4\height}{\includegraphics[width=0.20\textwidth]{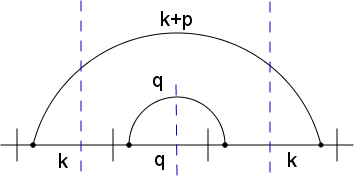}} = \notag \\ 
    - & \raisebox{-0.4\height}{\includegraphics[width=0.20\textwidth]{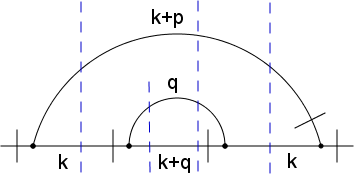}} + \raisebox{-0.4\height}{\includegraphics[width=0.20\textwidth]{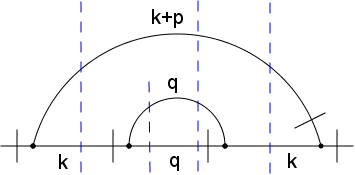}}\,.
\end{align}
The last line provides a compact representation of the same subtraction scheme: the first diagram corresponds to the combination  $I_H(1,1)-I_H(0,1)$, while the second corresponds to $-I_H(1,0)+I_H(0,0)$. In this form, the cancellation of the quadratic divergence becomes explicit, as both diagrams contain two time cuts crossing the subgraph, rendering them only logarithmically divergent; these remaining subdivergences cancel between the two terms.

As in the previous case \eqref{diag2_unrenorm}, to extract the poles of the original diagram H, one needs to obtain the poles of the subtraction terms (the last three diagrams in the second line  in \eqref{H_diag_renorm}). This task can be simplified by factorizing the subgraphs out of these terms. This can be achieved by factoring out external momenta and rescaling the integrals; for example, for the first term one finds:
\begin{equation}
    \raisebox{-0.4\height}{\includegraphics[width=0.12\textwidth]{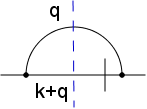}} = k^{2-2\varepsilon} \, \cdot \, \raisebox{-0.4\height}{\includegraphics[width=0.12\textwidth]{pics/diag_H_subgraph1.png}} \left.\vphantom{\rule{0pt}{15pt}}\right|_{k=1}.
\end{equation}
This allows one to rewrite the first subtraction term as
\begin{equation}
    \label{diagH_subtr1}
    I_H(0,1) = \raisebox{-0.4\height}{\includegraphics[width=0.2\textwidth]{pics/diag_H_sub01.png}} = \underbrace{\int d^{6-2\varepsilon}k\, \frac{1}{k^{2\varepsilon}\, (k+p)^2 \,\left[ k^2 + (k+p)^2 \right]^2}}_{I^{(1)}} \,\times\, \underbrace{\int d^{6-2\varepsilon}q\, \frac{1}{q^2 \,\left[ q^2 + (k+q)^2 \right]}}_{I^{(2)}}\left.\vphantom{\rule{0pt}{20pt}}\right|_{k=1}.
\end{equation}
Analogously, for the sum of the second and third terms,
\begin{align}
    \label{diagH_subtr2}
    &-I_H(1,0)+I_H(0,0) = \raisebox{-0.4\height}{\includegraphics[width=0.2\textwidth]{pics/diag_H_sub00-10.png}}  = \notag \\ & = \underbrace{\int d^{6-2\varepsilon}k\, \frac{1}{(p+k)^{2\varepsilon}\, k^2 \,\left[ k^2 + (k+p)^2 \right]^2}}_{I^{(1)}} \,\times\, \underbrace{\int d^{6-2\varepsilon}q\, \frac{1}{q^2 \, 2q^2 \, \left[ 2q^2 + (k+p)^2 \right]}}_{I^{(3)}}\left.\vphantom{\rule{0pt}{15pt}}\right|_{k+p=1}.
\end{align}
With a simple change of variables one can show that the first integrals in \eqref{diagH_subtr1} and \eqref{diagH_subtr2} are identical; we denote them as $I^{(1)}$. 

The integrals $I^{(1)}$ and $I^{(3)}$ are log-divergent, while $I^{(2)}$ is quadratically divergent. Rewriting the former using Feynman parametrization immediately isolates the required poles in the $\Gamma$-function prefactor; the remaining convergent integrals are evaluated with \textit{HyperInt}.  For $I^{(2)}$, one can differentiate it three times with respect to $k$ to make it convergent and thus suitable for direct evaluation. Applying the derivatives produces the pole:
\begin{equation}
    \label{H_diag_3_derivs}
    I^{(2)} = k^{2-2\varepsilon}\,I^{(2)}\Big|_{k=1} =  \frac{k^3}{(2-2\varepsilon)(1-2\varepsilon)(-2\varepsilon)} \,\partial_k \partial_k \partial_k I^{(2)}.
\end{equation}

Finally, we can specify the expression for extracting the poles of the diagram H at the point $p=1$ using the renormalized form \eqref{renorm_int} together with the introduced $\tilde{\mathcal{R}'}$-operation for the quadratic subgraph \eqref{H_diag_renorm_nodiag}:
\begin{equation}
    \partial_p I_H(1,1)\Big|_{p=1} =  \partial_p  \tilde{\mathcal{R}'}I_H(1,1)\Big|_{p=1} + \partial_p I_H(0,1)\Big|_{p=1} -  \partial_p \left[-I_H(1,0)+I_H(0,0)\right]\Big|_{p=1}\,.
\end{equation}
Since both the original diagram and its counterterms have a logarithmic superficial divergence and are therefore proportional to $p^{-4\varepsilon}$, differentiation with respect to $p$ simply produces a factor $-4\varepsilon$. Taking into account \eqref{diagH_subtr1},\eqref{diagH_subtr2}, and \eqref{H_diag_3_derivs}, we arrive at 
\begin{align}
    I_H(1,1)\Big|_{p=1} & = -\frac{1}{4\varepsilon} \partial_p  \tilde{\mathcal{R}'}I_H(1,1)\Big|_{p=1} + I_H(0,1)\Big|_{p=1} - \left[-I_H(1,0)+I_H(0,0)\right]\Big|_{p=1} = \notag \\
    & = -\frac{1}{4\varepsilon} \partial_p  \tilde{\mathcal{R}'}I_H(1,1)\Big|_{p=1} + \left[I^{(1)} \times \frac{1}{(2-2\varepsilon)(1-2\varepsilon)(-2\varepsilon)} \partial_k^3 I^{(2)}\Big|_{k=1} \right]\Bigg|_{p=1} - \left[I^{(1)} \times I^{(3)}\Big|_{k+p=1}\right]\Bigg|_{p=1}\,.
\end{align}
The integrals $\partial_p \tilde{\mathcal{R}'}I_H(1,1)$, $I^{(1)}$, $I^{(3)}$ and $\partial_k^3 I^{(2)}$ are then expressed using Feynman parametrization and evaluated with \textit{HyperInt}, which allows one to fully reconstruct the pole structure of diagram H.

\begin{comment}
\setlength{\extrarowheight}{4pt}
\begin{table}[ht]
\centering
\small
\begin{tabular}{c c c c}
\hline
№ & \normalfont \normalsize Nickel index & $\varepsilon^{-2}$ & $\varepsilon^{-1}$ \\
\hline
B & e12|e3|33||:0\_1\_0|0\_1|1\_1||:(1,2)|()|(2,)|() & $1/32$ & $31/192 - 11/16\,\ln(2)+9/32\,\ln(3)$ \\
C & e12|23|3|e|:0\_1\_1|1\_1|1|0|:()|()|()|(1,) & $1/16$ & $7/32 - \ln(2)/4$   \\
D & e12|23|3|e|:0\_0\_0|1\_1|1|0|:(1,2,3)|(2,3)|(3,)|() & $1/64$ & $31/384-11/32\,\ln(2)+9/64\,\ln(3)$   \\
E & e12|23|3|e|:0\_0\_1|0\_1|1|0|:(1,2,3)|(2,3)|(3,)|() & $1/64$ & $1/128+7/32\,\ln(2)-9/64\,\ln(3)$  \\
F & e12|23|3|e|:0\_0\_1|1\_1|1|0|:(1,2)|(2,)|()|() & -  & 1/24 \\
G & e12|e3|33||:0\_1\_1|0\_1|1\_1||:()|(1,)|()|() & $-1/48$ & $-25/288 + \ln(2)/12$ \\
H & e12|e3|33||:0\_1\_0|0\_1|0\_1||:(1,2,3)|()|(2,3)|(3,) & $-5/384$ & $-143/2304 + 37/192 \, \ln(2) - 9/128 \, \ln(3)$  \\
\hline
\end{tabular}
\caption{2 loop}
\label{tab:coeffs}
\end{table}
\end{comment}

The described procedure for restoring poles is applied for all one- and two-loop diagrams shown in Fig.~\ref{2loop_reduced}. The resulting coefficients of the $\varepsilon$ expansion are presented in Table~\ref{table_analyt}. Additional factors arising from symmetry and diagram reduction, together with the factor $1/2$ from the definition of the Green function considered $\Gamma_{\psi'\psi'}=\langle\psi'\psi'\rangle_{1\text{-irr}}/(2\lambda)$, are combined and denoted by $S$.
\setlength{\extrarowheight}{4pt}
\begin{table}[h!]
\centering
\small
\caption{Analytic coefficients of $\varepsilon$-expansion for one- and two-loop diagrams from Fig.~\ref{2loop_reduced}. $S$ denotes the overall combinatorial factor, including contributions from diagram symmetries, diagram reduction, and the factor $1/2$ arising from the definition of the considered Green function $\Gamma_{\psi' \psi'} = \langle \psi' \psi' \rangle_{1-\text{irr}} / (2 \lambda)$.} 
\begin{ruledtabular}
\begin{tabular}{c >{\scriptsize\ttfamily}c c c c}
%\begin{tabular}{|c | >{\scriptsize\ttfamily}c | c | c | c|}
Diagram & \normalfont \normalsize Nickel index & S & \multicolumn{2}{c}{ Analytic coefficients}\\
\hline
 &  &  & \multicolumn{2}{c}{\textbf{One loop}}\\
 \cline{4-5}
 &  &  & $\varepsilon^{-1}$ & $\varepsilon^{0}$ \\
\hline
A & e11|e|:0\_1\_1|0|:()|(1,)\footnote{The one-loop diagram has two identical time versions, which is taken into account in $S$.} & $1/2$ & $1/4$ & $3/8 - \ln(2)/2$ \\
\hline 
&  &  & \multicolumn{2}{c}{\textbf{Two loop}}\\
\cline{4-5}
 &  &  & $\varepsilon^{-2}$ & $\varepsilon^{-1}$ \\
\hline
B & e12|e3|33||:0\_1\_0|0\_1|1\_1||:(1,2)|()|(2,)|() & $1/2$ & $1/32$ & $31/192 - 11/16\,\ln(2)+9/32\,\ln(3)$ \\
C & e12|23|3|e|:0\_1\_1|1\_1|1|0|:()|()|()|(1,) & $1/2$ & $1/16$ & $7/32 - \ln(2)/4$   \\
D & e12|23|3|e|:0\_0\_0|1\_1|1|0|:(1,2,3)|(2,3)|(3,)|() & $1$ & $1/64$ & $31/384-11/32\,\ln(2)+9/64\,\ln(3)$   \\
E & e12|23|3|e|:0\_0\_1|0\_1|1|0|:(1,2,3)|(2,3)|(3,)|() & $1$ & $1/64$ & $1/128+7/32\,\ln(2)-9/64\,\ln(3)$  \\
F & e12|23|3|e|:0\_0\_1|1\_1|1|0|:(1,2)|(2,)|()|() & $1$ & \textemdash  & $1/24$ \\
G & e12|e3|33||:0\_1\_1|0\_1|1\_1||:()|(1,)|()|() & $1/2$ & $-1/48$ & $-25/288 + \ln(2)/12$ \\
H & e12|e3|33||:0\_1\_0|0\_1|0\_1||:(1,2,3)|()|(2,3)|(3,) & $1$ & $-5/384$ & $-143/2304 + 37/192 \, \ln(2) - 9/128 \, \ln(3)$  \\
\end{tabular}
\end{ruledtabular}
\label{table_analyt}
\end{table}

The first step toward the three-loop calculation is to determine the correction terms in the expansion of lower-loop diagrams: the $\varepsilon^0$ and $\varepsilon^1$ contributions for the one-loop diagram and the $\varepsilon^0$ contribution for the two-loop diagrams. We have carried out this computation with parametric integration; the resulting expressions are not presented due to their considerable length. The extension to three-loop order has not been completed because linearly irreducible three-loop diagrams arise, which preclude the straightforward application of the described method. The primary strategy to restore reducibility is an appropriate change of the integration variables \cite{Panzer_lr14, RatioRoots}; however, this procedure is not trivial, is not automated, and often depends on careful choices and intuition. Moreover, some integrals, although known to be expressible in terms of GPLs, cannot be evaluated using parametric integration. A detailed discussion of the restoration of the reducibility in dynamic diagrams is given in \cite{ADEK26}. Therefore, obtaining an analytic three-loop expression within this approach may be possible, although this task has yet to be solved.

\section{Three-loop RG functions}
\label{sec:6_Three-loop_RG_functions}
The three-loop calculation was carried out numerically using a custom implementation of the \textit{Sector Decomposition} algorithm adapted to the evaluation of dynamic diagrams. The results obtained for both two- and three-loop diagrams are provided in the Supplemental Material~\cite{SM}. We note that the numerical results for the two-loop diagrams shown in Fig.~\ref{2loop_reduced} are in full agreement with the analytic results listed in Table~\ref{table_analyt}. 

The contributions $A^{(l)}$ in \eqref{A_i} of the $l$-loop diagrams to the Green function considered \eqref{gamma_expansion} in the three-loop approximation are as follows:
\begin{align}
    \label{diag_contr}
    &A^{(1)} = \frac{1}{8\varepsilon} + \frac{3}{16} - \frac{\ln(2)}{4} + \left(\frac{3}{8} - \frac{\pi^2}{48} + \frac{\ln(2)}{8} - \frac{\ln^2(2)}{4} \right)\varepsilon\,, \notag \\
    &A^{(2)} = \frac{7}{128\varepsilon^2} +\left( \frac{55}{256} - \frac{23\ln(2)}{64} + \frac{9\ln(3)}{128}\right)\frac{1}{\varepsilon} + 0.10185185(12)\,, \notag \\
    %&A^{(2)}_{MV} = \frac{7}{128\varepsilon^2} +\left( \frac{55}{256} - \frac{23\ln(2)}{64} + \frac{9\ln(3)}{128}\right)\frac{1}{\varepsilon} + 0.10185184(12)\,, \notag \\
    &A^{(3)} = 0.02962240(3)\frac{1}{\varepsilon^3}+0.0625620(4)\frac{1}{\varepsilon^2}+0.1066921(3)\frac{1}{\varepsilon}\,.
    \notag \\
    %&A^{(3)}_{MV} = 0.02962360000\frac{1}{\varepsilon^3}+0.06256650000\frac{1}{\varepsilon^2}+0.1066921(5)\frac{1}{\varepsilon}\,.
\end{align}

Substituting these values into \eqref{gamma_expansion} enables the evaluation of the renormalization constant $Z_1$ \eqref{MS} from the requirement of the absence of poles in the renormalized Green function:
\begin{align}
    \label{Z1}
    Z_1(\varepsilon,u) = &\, 1
    + \Bigl(\frac{1}{8\varepsilon}\Bigr)u
    + \Bigl(\frac{7}{128\varepsilon^2} + \left( -\frac{13}{256}+\frac{9}{128} \ln\left(\frac{4}{3}\right) \right)\frac{1}{\varepsilon}\Bigr)u^2
    + \notag \\
    &  \Bigl(\frac{0.02962240(3)}{\varepsilon^3} - \frac{0.0552650(4)}{\varepsilon^2} + \frac{0.0312056(2)}{\varepsilon}\Bigr)u^3 + \mathcal{O}(u^4)\,.
\end{align}

%\begin{align}
%    Z_1MV(\varepsilon,u) = &\, 1
%    + u \Bigl(-\frac{1}{8\varepsilon}\Bigr)
%    + u^2 \Bigl(\frac{7}{128\varepsilon^2} + \left( -\frac{13}{256}+\frac{9\ln(2)}{64} - \frac{9\ln(3)}{128} \right)\frac{1}{\varepsilon}\Bigr)
%    + \notag \\
%    &  u^3 \Bigl(-\frac{0.02962360000}{\varepsilon^3} + \frac{0.0552604801}{\varepsilon^2} - \frac{0.0312056(5)}{\varepsilon}\Bigr) + \mathcal{O}(u^4)\,.
%\end{align}
The analytic coefficients up to the $u^2$ order, obtained with the hyperlogarithm integration, coincide with the result of \cite{Janssen1981}\footnote{Note that in \cite{Janssen1981} the authors used a different convention $d=6-\varepsilon$ instead of $d=6-2\varepsilon$}. 

The renormalization constant $Z_1$ corresponds to the anomalous dimension $\gamma_1$
\begin{equation}
    \label{gamma1_general}
    \gamma_1(u)=\beta(u)\partial_u\ln Z_1\,,
\end{equation}
where $\beta(u)$ is the RG beta-function. Due to the equivalence of the static and dynamic renormalization constant $Z_g$ \eqref{stat_dyn_equiv}, we employ the expression for the $\beta$-function from \cite{BGKS_phi321}, where it was calculated up to the five-loop order:
\begin{equation}
    \beta(\varepsilon,u)=-2{\varepsilon}u +\frac{3}{2}\,{u}^{2}-{\frac {125\,{u}^{3}}{72}}+ \left( {\frac{33085}{10368}}+\frac{5\,\zeta(3)}{4} \right) {u}^{4}+\ldots+\mathcal{O}(u^{7})\,.
\end{equation}
%\begin{equation}
%    \beta(\varepsilon,g)=-{\varepsilon}\,g +\frac{3}{4}\,{g}^{3}-{\frac {125\,{g}^{5}}{144}}+ \left( {\frac{33085}{20736}}+\frac{5\,\zeta(3)}{8} \right) {g}^{7}+\ldots+\mathcal{O}(g^{13})\,.
%\end{equation}
%We recall the coupling definition introduced in Section \ref{sec2: Renormalization of the model}, $u = (ig)^2\, S_d / (2\pi)^d$. 
In fact, because of the relation between the higher-order poles to the simple poles, $\gamma_1$ can be extracted using only the first poles in $Z_1$:
\begin{equation}
    \label{gamma1_simple}
    \gamma_1(u)=-2u\,\partial_u Z_1^{(1)} = -\frac{1}{4}u + \left( \frac{13}{64} - \frac{9}{32}\ln\left(\frac{4}{3}\right) \right)u^2 - 0.1872337(12)u^3 + \mathcal{O}(u^4)\,.
\end{equation}
%\begin{equation}
%    \label{gamma1_simple}
%    \gamma_1(g)=-g\,\partial_g Z_1^{(1)} = -\frac{1}{4}g^2 + \left( \frac{13}{64} - \frac{9}{32}\ln\left(\frac{4}{3}\right) \right)g^4 -0.1872337(12)g^6 + \mathcal{O}(g^8)\,.
%\end{equation}
%\begin{equation}
%    \label{gamma1_simple}
%    \gamma_1MV(g)=-g\,\partial_g Z_1^{(1)} = -\frac{1}{4}g^2 + \left( \frac{13}{64} - \frac{9}{32}\ln\left(\frac{4}{3}\right) \right)g^4 - 0.1872335533g^6 + \mathcal{O}(g^8)\,.
%\end{equation}
Since the RG functions are guaranteed to be finite, the requirement of the pole cancellation in \eqref{gamma1_general} can be exploited to reconstruct the analytic form of the third- and second-order poles in $A^{(3)}$ \eqref{diag_contr}
\begin{equation}
    A^{(3)} = \frac{91}{3072}\frac{1}{\varepsilon^3} + \left(\frac{5765}{27648} - \frac{43 \ln(2)}{128} + \frac{81 \ln(3)}{1024}\right) \frac{1}{\varepsilon^2} + 0.1066921(3)\frac{1}{\varepsilon}\,.
\end{equation}
This result leads to a refined expression for the $u^3$ coefficient in $Z_1$ \eqref{Z1}:
\begin{equation}
    Z_1^{(3)} = -\frac{91}{3072} \frac{1}{\varepsilon^3} + \left( \frac{3755}{55296} - \frac{45 \ln(2)}{512} + \frac{45 \ln(3)}{1024} \right) \frac{1}{\varepsilon^2} - 0.0312056(2) \frac{1}{\varepsilon} \,.
\end{equation}

The dynamic critical exponent $z$ is then determined as
\begin{equation}
    \label{exp_z}
    z = 2 - 2\gamma_{\phi}(u_*)+\gamma_1(u_*)\,,
\end{equation}
where $u_*$ is a fixed point defined by $\beta(u_*)=0$, and $\gamma_{\phi}$ is a field anomalous dimension. Both quantities are known from the static $\phi^3$ theory \cite{BGKS_phi321,KP21_phi3}:
\begin{align}
    \label{gzv}
    u_*(\varepsilon) = & \, \frac{4}{3}\varepsilon + \frac{500}{243}\varepsilon^2 + \left( \frac{25745}{19683} - \frac{160 \zeta(3)}{81}  \right)\varepsilon^3 + \left( - \frac{16 \pi^4}{729} + \frac{56512  \zeta(3)}{6561} - \frac{5120  \zeta(5)}{729} + \frac{7667410}{1594323} \right)\varepsilon^4 +  \notag \\ & \left( - \frac{640 \pi^{6}}{45927}
+ \frac{33038 \pi^{4}}{295245}
- \frac{18944 \zeta(3)^{2}}{729}
- \frac{23228768 \zeta(3)}{177147}
- \frac{6137440 \zeta(5)}{19683}
+ \frac{351680 \zeta(7)}{729}
+ \frac{1801355183}{516560652} \right)\varepsilon^5 +\mathcal{O}(\varepsilon^6),
\end{align}

\begin{align}
    \label{gamma_phi}
    \gamma_{\phi}(u) = & -\frac{1}{12}u + \frac{13}{432}u^2 + \left( -\frac{5195}{62208} + \frac{\zeta(3)}{24} \right)u^3 + \left(
\frac{53449}{248832}
+ \frac{7 \zeta(4)}{96}
+ \frac{35 \zeta(3)}{864}
- \frac{5 \zeta(5)}{18}
\right) u^4 + \notag \\ & \left(
- \frac{125 \zeta(6)}{288}
- \frac{5651 \zeta(4)}{27648}
- \frac{16492987}{20155392}
- \frac{25 \zeta(3)^2}{144}
- \frac{56693 \zeta(3)}{62208}
+ \frac{4471 \zeta(5)}{10368}
+ \frac{147 \zeta(7)}{64}
\right)u^5 +\mathcal{O}(u^6)\,.
\end{align}
\begin{comment}
The expression for the anomalous dimension $\gamma_1(u)$ \eqref{gamma1_simple} in the fixed point $u_*$ \eqref{gzv} is given by
\begin{equation}
    \label{gamma1_in_uzv}
    \gamma_1(u_*)=-\frac{1}{3}\varepsilon + \left( -\frac{149}{972} - \frac{1}{2}\ln\left( \frac{4}{3} \right) \right)\varepsilon^2 + 0.4933864(3) \varepsilon^3 + \mathcal{O}(\varepsilon^4)\,.
\end{equation}
\end{comment}
Substituting \eqref{gamma1_simple}, \eqref{gzv}, and \eqref{gamma_phi} into \eqref{exp_z}, we finally obtain the three-loop approximation for the exponent $z$:
\begin{equation}
    \label{z_final}
    z = 2-\frac{1}{9}\varepsilon+\left( \frac{241}{2916} - \frac{1}{2}\ln\left(\frac{4}{3}\right) \right)\varepsilon^2 + 0.143867(3)\,\varepsilon^3 + \mathcal{O}(\varepsilon^4)\,.
\end{equation}
%\begin{equation}
%    z_{MV} = 2-\frac{1}{9}\varepsilon+\left( \frac{241}{2916} - \frac{1}{2}\ln \left(\frac{4}{3} \right) \right)\varepsilon^2 + 0.1454(15)\,\varepsilon^3 + \mathcal{O}(\varepsilon^4)\,.
%\end{equation}

\section{Resummation}
\label{sec:7_Resummation}
Perturbative series for critical exponents obtained within the renormalization-group (RG) framework are asymptotic: their coefficients exhibit factorial growth, which precludes direct extraction of reliable estimates comparable to experimental or numerical results. In addition, the expansion parameter $\varepsilon$, formally assumed small, exceeds unity in the physically relevant dimensions; hence a resummation procedure is required. Numerous resummation strategies exist, and their accuracy is strongly dependent on the number of available perturbative coefficients. More advanced techniques incorporate additional information, for example, high-order asymptotics or exact values known in specific dimensions. For the resummation of the three-loop expression \eqref{z_final} for the exponent $z$, we employ standard Padé approximants and two Borel-based procedures that suppress factorial growth: the Padé-Borel-Leroy method and the KP17 approach \cite{KP17}.

\begin{comment}
Two possible ways exist to obtain the resummed value of the dynamical exponent $z$. One can resum directly the expansion \eqref{z_final} for $z$\footnote{In practice we resum $z-2$; this choice is immaterial.}, or resum the perturbative series for the anomalous dimension $\gamma_1$ given in \eqref{gamma1_in_uzv}. In the latter approach the relation \eqref{exp_z} is used, with the resummed value of the static exponent $\eta$ taken from \cite{BGKS_phi321,Gracey_sixloop_YLES}. Since neither approach is theoretically preferred, the final estimate is defined as the arithmetic mean of the two results.
\end{comment}

The Padé approximant is defined as the rational function
\begin{equation}
    P_{\{L/M\}}(\varepsilon)=\frac{P_L(\varepsilon)}{P_M(\varepsilon)}\,,
\end{equation}
where $P_L(\varepsilon)$ and $P_M(\varepsilon)$ are polynomials of order $L$ and $M$, respectively. Their coefficients are chosen so that the expansion of the approximant up to $L+M+1$ terms coincides with those of the original series. Different approximants can produce substantially different estimates, raising the problem of selecting the appropriate ones. We follow the strategy of \cite{AIKKS19} and discard approximants with poles in the vicinity of $\varepsilon_{\text{phys}}$ as well as maximally off-diagonal forms [0/M] and [L/0], since these are known to perform poorly. For the considered series \eqref{z_final} only two approximants survive, [2/1] and [1/2], because [1/1] possesses a pole.

The Padé-Borel-Leroy (PBL) resummation procedure employs the Borel-Leroy transformation to compensate for the factorial growth of the series coefficients, thereby accelerating convergence and yielding more precise estimates:
\begin{equation}
    \label{PBL}
    f(x) = \sum_{i=0}^{\infty} c_i x^i 
     = \int_{0}^{\infty} e^{-t} t^b F(x t) \, dt, 
\quad 
F(y) = \sum_{i=0}^{\infty} \frac{c_i}{\Gamma(i+b+1)} y^i\,.
\end{equation}
The Borel transform $F(y)$ is approximated by Padé approximants subject to the same selection criteria described above, except that approximants with poles on the positive real axis are now excluded, rather than only those with poles in the vicinity of the resummation point. Substituting the approximant into the integral and performing the integration yields the resummed value. As for the ordinary Padé resummation, the only suitable approximants in this case are [2/1] and [1/2]. The parameter $b$ in \eqref{PBL} is usually tuned to minimize discrepancies between different Padé approximants. To determine its optimal value, we perform a discrete scan over $b \in [0, 20]$ \cite{AIKKS19}. The optimum is found at the upper boundary of this interval, since the two approximants approach each other slowly as $b$ increases. Although higher values of $b$ would further reduce the discrepancy, they would also suppress the contribution of the higher-order coefficients in the series. We therefore adopt $b_{\text{opt}}=20$ as a compromise. Moreover, variations of $b$ in this range have almost no effect on the final resummed value.

The final resummation method employed is KP17, introduced in a six-loop study of the static $\phi^4$ theory \cite{KP17}. It is an advanced variant of Borel summation with conformal mapping that fine-tunes multiple fitting parameters according to a fastest-convergence criterion. Its accuracy has strong dependence on the number of known perturbative coefficients; therefore, high precision should not be expected when only three coefficients are available. Since KP17 involves several technical steps and contains many practical subtleties, we do not reproduce its description here and refer the reader to the original paper for details \cite{KP17}.

The difficulty of assigning a reliable error bound to resummation results is well known. Here, we adopt the uncertainty-estimation procedure of \cite{AIKKS19}; all error bars in this section are computed according to that prescription. The estimates of the individual Padé and PBL approximants are treated as independent measurements, and the final estimate $x^{(i)}$ for each method $i$ is taken as their arithmetic mean. The corresponding error bars $\Delta x^{(i)}$ are then defined using the Student t-distribution $t_{p,n}$ at confidence level $p=0.95$:  
\begin{equation}
    x^{(i)} = \frac{x_1 + \dots + x_n}{n},
    \qquad
    \Delta x^{(i)} = t_{0.95,n}\sqrt{\frac{(x^{(i)} - x_1)^2 + \dots + (x^{(i)} - x_n)^2}{n(n-1)}}\,.
\end{equation}
The uncertainty for the KP17 estimates is determined by the internal criterion of the method. The final estimate is obtained as a weighted average of all results of the different methods $x^{(i)}$, with weights $w_i$ proportional to the inverse uncertainties $1/\Delta x^{(i)}$:
\begin{equation}
    x_{\text{final}} = \frac{\sum_i x^{(i)} w_i}{\sum_i w_i}\,.
\end{equation}
The final uncertainty is computed by combining two contributions: the weighted standard deviation and the weighted average of the individual uncertainties
\begin{equation}
    E_{\text{final}}=\sqrt{E_1^2+E_2^2}\,,\,\,\,\,\,\,\text{where}\quad  E_1=\sqrt{\frac{\sum_i(x_{\text{final}}-x^{(i)})^2 w_i}{\sum_i w_i}}\,, \quad E_2=\frac{1}{\sum_i w_i}\,.
\end{equation}

All resummation methods discussed above (Padé, PBL, and KP17) are applied to the series \eqref{z_final} for various values of $d$; the resulting estimates are reported in Table~\ref{z_resum_table}. Uncertainties in the third-order coefficients, arising from their numerical evaluation, contribute negligibly to the resummation results compared with the uncertainties introduced by the resummation procedures themselves; accordingly, we use the central values in all calculations. It is evident that the KP17 results deviate substantially from the Padé and PBL estimates. This discrepancy can be attributed to the limited number of known coefficients in the original series and the absence of additional information on its high-order behavior. For this reason, we exclude the KP17 outcomes and use only the Padé and PBL estimates in the final determination of the exponent $z$.

\begin{table}[h]
    \centering
    \caption{Resummed estimates of the exponent $z$ for different spatial dimensions $d$. The final results are based on Pad\'e and PBL estimates, excluding KP17. For the Pad\'e and Pad\'e-Borel-Leroy (PBL) methods, the only admissible approximants [2/1] and [1/2] are used. For the PBL method, the tuning parameter is set to $b=20$. Previous results from the two-loop calculation \cite{Zhong2017} and functional RG \cite{Zhong_FRG} are included for comparison.}
    \begin{ruledtabular}
    \begin{tabular}{lccccc}
        & $d=1$ & $d=2$ & $d=3$ & $d=4$ & $d=5$ \\
        \midrule
        Pad\'e & 1.69(8) & 1.75(5) & 1.81(3) & 1.873(9)  & 1.9378(15) \\
        PBL    & 1.68(8) & 1.75(5) & 1.81(3) & 1.873(10) & 1.9377(15) \\
        KP17   & 1.7(4)  & 1.8(3)  & 1.8(2)  & 1.87(9)   & 1.94(3) \\
        \midrule
        Final $z$ (KP17 excluded) & 1.69(4) & 1.75(3) & 1.808(15) & 1.873(5) & 1.9378(8) \\
        \midrule
        Two-loop RG \cite{Zhong2017} & 1.677 & 1.753(10) & 1.817(8) & 1.880(6) & 1.941(3) \\
        FRG \cite{Zhong_FRG}      & --    & --        & 1.661    & 1.809    & 1.922 \\
    \end{tabular}
    \end{ruledtabular}
    \label{z_resum_table}
\end{table}

The final results are consistent with previous resummed two-loop estimates \cite{Janssen1981, Zhong2017} and are reasonably close to the functional RG result \cite{Zhong_FRG}, for which no error estimate was provided. However, it should again be emphasized that the availability of only three coefficients in the perturbative expansion severely limits the possibility of assigning a well-justified resummation uncertainty.

\section{Summary}
\label{sec:8_Summary}
We have calculated the dynamic critical exponent $z$ of the Yang--Lee edge singularity in relaxational (model A) dynamics using perturbative renormalization group to three-loop order. The calculation combined a diagram reduction algorithm, an analytic two-loop evaluation via parametric integration with hyperlogarithms, and a Sector Decomposition implementation for the numerical evaluation of three-loop integrals. The analytical stage used a recently developed framework for applying parametric integration to multiloop calculations in models of critical dynamics \cite{AEKTurb2024, ADEK26}. This approach relies on a specific regularization scheme for dynamic diagrams that enables the use of the hyperlogarithm framework developed by F. Brown and E. Panzer \cite{Brown08, Panzer15}. Our two-loop result reproduces the previously known analytic expression \cite{Janssen1981}. An extension to higher orders requires further manipulations to address the linear irreducibility of certain three-loop diagrams; this remains to be completed. The three-loop contribution was obtained numerically using a custom version of the Sector Decomposition algorithm adapted for dynamic diagrams. We resummed the perturbative series using Padé, Padé-Borel-Leroy, and KP17 \cite{KP17} methods in various spatial dimensions. Since KP17 is known to produce large uncertainties when only a few coefficients of expansion are available, our final estimates are based exclusively on the results from Padé and Padé-Borel-Leroy. The values obtained are in good agreement with previous two-loop estimates \cite{Janssen1981, Zhong2017} and with the functional RG calculation \cite{Zhong_FRG}.

\section*{Acknowledgments}
We are grateful to F.~Zhong for drawing our attention to this problem and for his valuable comments. D.A.D., D.A.E., and M.V.K.  gratefully acknowledge the support of Foundation for the Advancement of Theoretical Physics ``BASIS" through Grant 25-1-2-48-1.
The work of D.A.D. was carried out at the Saint Petersburg Leonhard Euler International Mathematical Institute and was supported by the Ministry of Science and Higher Education of the Russian Federation (agreement no. 075–15–2025–343). We would also like to thank E. Zerner-Käning for careful reading and editing.

\section*{Data Availability}
The complete lists of the reduced two- and three-loop time versions, together with the corresponding numerical evaluation results, are provided in the Supplemental Material~\cite{SM}. The implementation of the Sector Decomposition method used for the numerical evaluation of the Feynman integrals is publicly available in \cite{SD_gitlab}.

\bibliography{main}

\end{document}

%% file: tikz_pics/rule4.tex
\begin{tikzpicture}[baseline=(current bounding box.center), node distance=1.2cm]

\coordinate[vertexBlue] (y1);
\coordinate [below=of y1] (y2);

\coordinate[vertexBlue] (v1);
\coordinate[right=of v1] (v2);
\coordinate (e1) at ($ (v2)!1.8!(v1) +  (y1)!0.6!(y2)$);
\coordinate (e2) at ($ (v2)!1.8!(v1) +  (y1)!0.3!(y2)$);
\coordinate[smallPoint] (e3) at ($ (v2)!1.7!(v1) +  (y1)!0.!(y2)$);
\coordinate[smallPoint] (e4) at ($ (v2)!1.7!(v1) +  (y1)!0.1!(y2)$);
\coordinate[smallPoint] (e5) at ($ (v2)!1.7!(v1) -  (y1)!0.1!(y2)$);
\coordinate (e6) at ($ (v2)!1.8!(v1) -  (y1)!0.3!(y2)$);
\coordinate (e7) at ($ (v2)!1.8!(v1) -  (y1)!0.6!(y2)$);

\dynamicLine{0.5} (e7) -- (v1);
\arcl{v1}{e2}{0};
\arcl{v1}{e6}{0};
\arcl{v1}{e1}{0};
%\draw[loosely dotted] (v1)--(e3) node {text};
%\path[draw=none] ($ (e3)!0.5!(v1)$) node {$\cdots$};
%\draw[densely dashed] (v1)--(e4);
%\draw[densely dashed] (v1)--(e5);

\coordinate (i1) at ($ (v1)!0.4!(v2) +  (y1)!0.17!(y2)$);
\coordinate[smallPoint] (i4) at ($ (v1)!0.35!(v2) +  (y1)!0.07!(y2)$);
\coordinate[smallPoint] (i2) at ($ (v1)!0.35!(v2) +  (y1)!0.!(y2)$);
\coordinate[smallPoint] (i5) at ($ (v1)!0.35!(v2) -  (y1)!0.07!(y2)$);
\coordinate (i3) at ($ (v1)!0.4!(v2) -  (y1)!0.17!(y2)$);

\draw (v1)--(i1);
%\draw[loosely dotted] (v1)--(i2);
\draw (v1)--(i3);

\coordinate (s1) at ($(e3)!.2!(v1) + (y1)!0.7!(y2)$);
\coordinate (s2) at ($(e3)!.2!(v1) - (y1)!0.7!(y2)$);

\timeVersion{s1}{s2};

\end{tikzpicture}

%% file: tikz_pics/rule3.tex
\begin{tikzpicture}[baseline=(current bounding box.center), node distance=1.2cm]

\coordinate (y1);
\coordinate [below=of y1] (y2);

\coordinate[vertexBlue] (v1);
\coordinate[right=of v1] (v2);
\coordinate (e1) at ($ (v2)!1.8!(v1) +  (y1)!0.6!(y2)$);
\coordinate (e2) at ($ (v2)!1.8!(v1) +  (y1)!0.3!(y2)$);
\coordinate[smallPoint] (e3) at ($ (v2)!1.7!(v1) +  (y1)!0.!(y2)$);
\coordinate[smallPoint] (e4) at ($ (v2)!1.7!(v1) +  (y1)!0.1!(y2)$);
\coordinate[smallPoint] (e5) at ($ (v2)!1.7!(v1) -  (y1)!0.1!(y2)$);
\coordinate (e6) at ($ (v2)!1.8!(v1) -  (y1)!0.3!(y2)$);
\coordinate (e7) at ($ (v2)!1.8!(v1) -  (y1)!0.6!(y2)$);

\dynamicLine{0.5} (e6) -- (v1);
\arcl{v1}{e2}{0};
\arcl{v1}{e1}{0};
\arcl{v1}{e7}{0};
%\draw[loosely dotted] (v1)--(e3);
%\path[draw=none] ($ (e3)!0.!(v1)$) node {\rotatebox{90}{$\dots$}};
%\draw[densely dashed] (v1)--(e4);
%\draw[densely dashed] (v1)--(e5);

\coordinate (i1) at ($ (v1)!0.4!(v2) +  (y1)!0.17!(y2)$);
\coordinate[smallPoint] (i4) at ($ (v1)!0.35!(v2) +  (y1)!0.07!(y2)$);
\coordinate[smallPoint] (i2) at ($ (v1)!0.35!(v2) +  (y1)!0.!(y2)$);
\coordinate[smallPoint] (i5) at ($ (v1)!0.35!(v2) -  (y1)!0.07!(y2)$);
\coordinate (i3) at ($ (v1)!0.4!(v2) -  (y1)!0.17!(y2)$);

\draw (v1)--(i1);
%\draw[loosely dotted] (v1)--(i2);
\draw (v1)--(i3);

\coordinate (s1) at ($(e3)!.2!(v1) + (y1)!0.7!(y2)$);
\coordinate (s2) at ($(e3)!.2!(v1) - (y1)!0.7!(y2)$);

\timeVersion{s1}{s2};

\end{tikzpicture}

%% file: tikz_pics/rule2.tex
\begin{tikzpicture}[baseline=(current bounding box.center), node distance=1.2cm]

\coordinate (y1);
\coordinate [below=of y1] (y2);

\coordinate[vertexBlue] (v1);
\coordinate[right=of v1] (v2);
\coordinate (e1) at ($ (v2)!1.8!(v1) +  (y1)!0.6!(y2)$);
\coordinate (e2) at ($ (v2)!1.8!(v1) +  (y1)!0.3!(y2)$);
\coordinate[smallPoint] (e3) at ($ (v2)!1.7!(v1) +  (y1)!0.!(y2)$);
\coordinate[smallPoint] (e4) at ($ (v2)!1.7!(v1) +  (y1)!0.1!(y2)$);
\coordinate[smallPoint] (e5) at ($ (v2)!1.7!(v1) -  (y1)!0.1!(y2)$);
\coordinate (e6) at ($ (v2)!1.8!(v1) -  (y1)!0.3!(y2)$);
\coordinate (e7) at ($ (v2)!1.8!(v1) -  (y1)!0.6!(y2)$);

\dynamicLine{0.5} (e2) -- (v1);
\arcl{v1}{e1}{0};
\arcl{v1}{e6}{0};
\arcl{v1}{e7}{0};
%\draw[densely dashed] (v1)--(e3);
%\draw[densely dashed] (v1)--(e4);
%\draw[densely dashed] (v1)--(e5);

\coordinate (i1) at ($ (v1)!0.4!(v2) +  (y1)!0.17!(y2)$);
\coordinate[smallPoint] (i4) at ($ (v1)!0.35!(v2) +  (y1)!0.07!(y2)$);
\coordinate[smallPoint] (i2) at ($ (v1)!0.35!(v2) +  (y1)!0.!(y2)$);
\coordinate[smallPoint] (i5) at ($ (v1)!0.35!(v2) -  (y1)!0.07!(y2)$);
\coordinate (i3) at ($ (v1)!0.4!(v2) -  (y1)!0.17!(y2)$);

\draw (v1)--(i1);
%\draw[loosely dotted] (v1)--(i2);
\draw (v1)--(i3);

\coordinate (s1) at ($(e3)!.2!(v1) + (y1)!0.7!(y2)$);
\coordinate (s2) at ($(e3)!.2!(v1) - (y1)!0.7!(y2)$);

\timeVersion{s1}{s2};

\end{tikzpicture}

%% file: tikz_pics/rule1.tex
\begin{tikzpicture}[baseline=(current bounding box.center), node distance=1.2cm]

\coordinate (y1);
\coordinate [below=of y1] (y2);

\coordinate[vertexBlue] (v1);
\coordinate[right=of v1] (v2);
\coordinate (e1) at ($ (v2)!1.8!(v1) +  (y1)!0.6!(y2)$);
\coordinate (e2) at ($ (v2)!1.8!(v1) +  (y1)!0.3!(y2)$);
\coordinate[smallPoint] (e3) at ($ (v2)!1.7!(v1) +  (y1)!0.!(y2)$);
\coordinate[smallPoint] (e4) at ($ (v2)!1.7!(v1) +  (y1)!0.1!(y2)$);
\coordinate[smallPoint] (e5) at ($ (v2)!1.7!(v1) -  (y1)!0.1!(y2)$);
\coordinate (e6) at ($ (v2)!1.8!(v1) -  (y1)!0.3!(y2)$);
\coordinate (e7) at ($ (v2)!1.8!(v1) -  (y1)!0.6!(y2)$);

\dynamicLine{0.5} (e1) -- (v1);
\arcl{v1}{e2}{0};
\arcl{v1}{e6}{0};
\arcl{v1}{e7}{0};
%\draw[densely dashed] (v1)--(e3);
%\draw[densely dashed] (v1)--(e4);
%\draw[densely dashed] (v1)--(e5);

\coordinate (i1) at ($ (v1)!0.4!(v2) +  (y1)!0.17!(y2)$);
\coordinate[smallPoint] (i4) at ($ (v1)!0.35!(v2) +  (y1)!0.07!(y2)$);
\coordinate[smallPoint] (i2) at ($ (v1)!0.35!(v2) +  (y1)!0.!(y2)$);
\coordinate[smallPoint] (i5) at ($ (v1)!0.35!(v2) -  (y1)!0.07!(y2)$);
\coordinate (i3) at ($ (v1)!0.4!(v2) -  (y1)!0.17!(y2)$);

\draw (v1)--(i1);
%\draw[loosely dotted] (v1)--(i2);
\draw (v1)--(i3);

\coordinate (s1) at ($(e3)!.2!(v1) + (y1)!0.7!(y2)$);
\coordinate (s2) at ($(e3)!.2!(v1) - (y1)!0.7!(y2)$);

\timeVersion{s1}{s2};

\end{tikzpicture}

%% file: tikz_pics/rule5.tex
\begin{tikzpicture}[baseline=(current bounding box.center), node distance=1.2cm]

\coordinate (y1);
\coordinate [below=of y1] (y2);

\coordinate[vertexBlue] (v1);
\coordinate[right=of v1] (v2);
\coordinate (e1) at ($ (v2)!1.8!(v1) +  (y1)!0.6!(y2)$);
\coordinate (e2) at ($ (v2)!1.8!(v1) +  (y1)!0.3!(y2)$);
\coordinate[smallPoint] (e3) at ($ (v2)!1.7!(v1) +  (y1)!0.!(y2)$);
\coordinate[smallPoint] (e4) at ($ (v2)!1.7!(v1) +  (y1)!0.1!(y2)$);
\coordinate[smallPoint] (e5) at ($ (v2)!1.7!(v1) -  (y1)!0.1!(y2)$);
\coordinate (e6) at ($ (v2)!1.8!(v1) -  (y1)!0.3!(y2)$);
\coordinate (e7) at ($ (v2)!1.8!(v1) -  (y1)!0.6!(y2)$);

%\dynamicLine{0.6} (e1) -- (v1);
\arcl{v1}{e1}{0};
\arcl{v1}{e2}{0};
\arcl{v1}{e6}{0};
\arcl{v1}{e7}{0};
%\draw[densely dashed] (v1)--(e3);
%\draw[densely dashed] (v1)--(e4);
%\draw[densely dashed] (v1)--(e5);

\coordinate (i1) at ($ (v1)!0.4!(v2) +  (y1)!0.17!(y2)$);
\coordinate[smallPoint] (i4) at ($ (v1)!0.35!(v2) +  (y1)!0.07!(y2)$);
\coordinate[smallPoint] (i2) at ($ (v1)!0.35!(v2) +  (y1)!0.!(y2)$);
\coordinate[smallPoint] (i5) at ($ (v1)!0.35!(v2) -  (y1)!0.07!(y2)$);
\coordinate (i3) at ($ (v1)!0.4!(v2) -  (y1)!0.17!(y2)$);

\draw (v1)--(i1);
%\draw[loosely dotted] (v1)--(i2);
\draw (v1)--(i3);

\coordinate (s1) at ($(e3)!.2!(v1) + (y1)!0.7!(y2)$);
\coordinate (s2) at ($(e3)!.2!(v1) - (y1)!0.7!(y2)$);

\draw[opacity=0.0, densely dashed, blue] (s1) -- (s2);

\end{tikzpicture}